\documentclass[11pt,twoside,A4]{article} 
\usepackage{times,fancyhdr}
\usepackage{graphicx}
\usepackage{latexsym} 
\usepackage[affil-it]{authblk}
\usepackage{mathptm}
\usepackage{mathtools}
\usepackage{amssymb}
\usepackage{color}
\usepackage{setspace}
\usepackage{hyperref}

\usepackage[top=4cm, bottom=4cm, left=4cm, right=4cm]{geometry}

\def\beq{\begin{equation}}
\def\eeq{\end{equation}}
\def\beqn{\begin{eqnarray}}
\def\eeqn{\end{eqnarray}}

\newcommand{\intsum}{
\hspace*{1.0mm} {\mbox{\Large$\Sigma$}}\hspace*{-3.8mm}\int}
\newcommand{\be}{\begin{equation}}
\newcommand{\ee}{\end{equation}}
\newcommand{\bea}{\begin{eqnarray}}
\newcommand{\eea}{\end{eqnarray}}

\begin{document}
 
\title{A Future-Input Dependent Path Integral for Quantum Mechanics}
\author{S.~Donadi, S.~Hossenfelder} 
\affil{\small Frankfurt Institute for Advanced Studies\\
Ruth-Moufang-Str. 1,
D-60438 Frankfurt am Main, Germany
}
\date{}
\maketitle
\vspace*{-1cm}
\begin{abstract}
We here put forward a new path-integral over Hilbert space and show that it reproduces quantum mechanics exactly. This approach works by optimizing the generating functional under a variation of the final state; it is hence an example of a future-input dependent theory. The benefit of this method is that entangled states appear in the paths directly, and the generating functional can be modified to also include wave-function collapse. 
\end{abstract}

\section{Introduction}

The standard way to use Feynman's path integral in quantum mechanics \cite{Feynman} is to sum over all possible paths in space-time, for example to calculate propagators or transition amplitudes. These paths are classical paths -- they have a well-defined position -- but they need not be classically allowed, that is, they need not solve the classical equation of motion. 

In the path integral we will consider here, in contrast, we do not sum over paths in space-time. Instead, we sum over paths in Hilbert space. These paths are in general not only not classically allowed, they are not classical to begin with. Each path corresponds to a possible evolution of the state $|\Psi(t)\rangle$, but only one of them solves the Schr\"odinger equation. The path integral we are about to construct thus contains paths that do not fulfill the Schr\"odinger equation in much the same way that the usual path integral contains paths that do not fulfill the Euler-Lagrange equations.

The motivation we have for considering this approach is three-fold. First, new mathematical methods can offer a different perspective on old problems, so we believe this new approach to quantum mechanics to be of general interest. Second, in the usual path integral, entangled states appear in the initial and final state, but have no natural interpretation on the paths themselves.
Our approach remedies this issue because we integrate over Hilbert space which includes entangled states. An attempt to use paths in Hilbert space instead was previously proposed in \cite{Green:2016gpb}, and a space-time based solution was recently discussed in \cite{WharthonPath}. Third, it is an example of a future-input-dependent formalism \cite{Wharton:2020qrq} and is hence a starting point to solve the measurement problem, as laid out in \cite{toymodel,Jonte}.

\section{The Sum over States}

The framework we use is the normal formalism of quantum mechanics: the state of the system is described by a vector, $|\Psi(t) \rangle$ of an Hilbert space $\mathcal{H}$ and $ H$ is the Hamiltonian operator of the system.
We then postulate that the dynamics of the system with initial state $|\Psi_{\rm i} \rangle$  will evolve to the final state $|\Psi_{\rm e} \rangle$  for which $| Z ( \Psi_{\rm i} \to \Psi_{\rm e} ) | $ is
 maximal, where
 \beqn
Z ( \Psi_{\rm i} \to \Psi_{\rm e} )  = \intsum_{|\Phi\rangle} \exp \left(  - \frac{{i}}{\hbar} \int_{t_{\rm i}}^{t_{\rm e}} {\rm d} t \langle \Phi (t) | {i}\hbar \partial_t - {H} | \Phi (t) \rangle  \right)~. \label{Z}
\eeqn
Here, $t_{\rm i}$ and $t_{\rm e}$ are the initial and final times, $\langle \cdot | \cdot \rangle$ denotes the usual inner product on the Hilbert space, and the integral is taken over all paths $|\Phi (t) \rangle$ that are time-sequences of vectors in the Hilbert space with initial
 value $|\Phi ({t_{\rm i})} \rangle = |\Psi_{\rm i}\rangle$ and end value $|\Phi(t_{\rm e}) \rangle =  |\Psi_{\rm e} \rangle$.

In the next section, we will show that this path integral reproduces quantum mechanics, that is, the end-state for which $| Z ( \Psi_{\rm i} \to \Psi_{\rm e} ) | $ is
 maximal is that which one also gets from the Schr\"odinger evolution. But before we do that, let us highlight the differences to the usual path integral of quantum mechanics:
 
 \begin{enumerate}
 \item The paths under consideration here are not trajectories in space-time but trajectories in  Hilbert-space. They are not only not classically allowed, they are not classical. 
 \item
The system is not only allowed to take all possible paths to one fixed final state, but can take all possible paths to all possible final states. This is why the model has a future-input dependence.
 \item $|Z|^2$ is {\sl not} a probability. As with the classical principle of least action, the probability is $=1$
 that the initial state evolves into the state which maximizes the transition function,
 and $=0$ that it goes into any other state. $|Z|^2$ is instead the quantity that is maximized. 
\item
 The paths that we integrate over need not solve the Schr\"odinger-equation, just like, in the normal path integral, the paths that one integrates over do not have to solve the classical Euler-Lagrange equations.
 \end{enumerate}
 
 \section{Calculation} \label{calc}
 
In order to define the sum over the states in Eq. (\ref{Z}), we first expand the wave-function in energy eigenstates $|E_{j}\rangle$
\begin{equation}
|\Phi(t)\rangle=\sum_{j}a_{j}(t)|E_{j}\rangle~.\label{expansion}
\end{equation}
With that, we define the sum in Eq.\ (\ref{Z}) as a sum over all possible values the coefficients $a_{j}(t)$ can take: 
\begin{equation}\label{Zexp}
Z( \Psi_{\rm i} \to \Psi_{\rm e} ):=\left( \prod_{j} \int_{a_{j}(t_{\rm i}),t_{\rm i}}^{a_{j}(t_{\rm e}),t_{\rm e}}D[a_{j}(t)]D[\bar{a}_{j}(t)]\right)\,e^{\frac{i}{\hbar} \int_{t_{\rm i}}^{t_{\rm e}}dt  \sum_j L_{j}[a_{j}(t),\bar a_{j}(t)]}~,
\end{equation}
where
\begin{equation}\label{Lj}
L_{j}[a_{j}(t),\bar a_{j}(t)]:=\frac{i\hbar}{2}\left(\bar{a}_{j}(t)\dot{a}_{j}(t)-\dot{\bar{a}}_{j}(t)a_{j}(t)\right)-E_{j}|a_{j}(t)|^{2}~,
\end{equation}
the bar denotes complex conjugation and the dot the time derivative.
In Eq. (\ref{Zexp}) the integral over $D[a_{j}(t)]D[\bar{a}_{j}(t)]$ denotes the standard path integration for complex variables. 
We have chosen to express the path integral using energy eigenstates just because it simplifies 
the following calculation. The path integral itself however does not depend on the choise of basis: Given another expansion $|\Phi(t)\rangle=\sum_{j}b_{j}(t)|B_{j}\rangle$ over an arbitrary basis $| B_j \rangle$, the Jacobian determinant for the change of variables from $a_j(t)$ to $b_j(t)$ is equal to 1 because the transformation is unitary. The integrand is likewise unitarily invariant.

The definitions above strictly speaking apply only when the energy-spectrum is discrete. The generalization to a continuous spectrum is possible but leads to technical difficulties which we believe to be neither relevant nor insightful for the purpose of this present discussion. We will therefore limit the following analysis to the case of a discrete energy spectrum.

The path integral in Eq. (\ref{Zexp}) can be written as a product of identical factors for each of the energy-eigenvalues 
\begin{equation}
Z( \Psi_{\rm i} \to \Psi_{\rm e} )=\prod_{j}Z_{j}( \Psi_{\rm i} \to \Psi_{\rm e} )~,
\end{equation}
where
\begin{equation}\label{Zj}
Z_{j}( \Psi_{\rm i} \to \Psi_{\rm e} ):=\int_{a_{j}(t_{\rm i}),t_{\rm i}}^{a_{j}(t_{\rm e}),t_{\rm e}}D[a_{j}(t)]D[\bar{a}_{j}(t)]\,e^{\frac{i}{\hbar}\int_{t_{\rm i}}^{t_{\rm e}}dtL_{j}[a_{j}(t),\bar a_{j}(t)]}~,
\end{equation}
and $L_{j}[a_{j}(t),\bar a_{j}(t)]$ is defined in Eq. (\ref{Lj}). It is because of this factorization that the energy-eigenbasis simplifies the calculation.

The path integral in Eq. (\ref{Zj}) could then be  computed directly by discretizing the time-sequence and performing the resulting integrals, which are Gaussian integrals given the form of $L_{j}$. However, there is a simpler way to do it which also reveals an interesting connection to a path integral appearing in a completely different context: the path integral for computing the propagator between coherent states. We will here follow the treatment laid out in more detail in \cite{Schulman} (chapter 27).

Given the Hilbert space of a harmonic oscillator with ladder operators $a$ and $a^\dagger$, coherent states are defined as $|z\rangle:=e^{za^{\dagger}-\bar{z}a}|0\rangle$, where $z$ is a complex number. The propagator between two coherent states $|z_{0}\rangle$ and $|z_{f}\rangle$ is by definition
\begin{equation}\label{cs_prop}
K(z_{f},z_{0},t):=\langle z_{f}|e^{-\frac{i}{\hbar}Ht}|z_{0}\rangle~,
\end{equation}
and it can be written in the form of a path integral as \cite{Schulman}:
\begin{flalign} \label{cs_path}
 & K(z_{f},z_{0},t) = \nonumber \\
&= \underset{N\rightarrow\infty}{\lim}\int\prod_{j=1}^{N-1}\frac{d^{2}z_{j}}{\pi}\exp\left(\sum_{j=0}^{N-1} \left[ \frac{1}{2}(\bar{z}_{j+1}-\bar{z}_{j})z_{j}-\frac{1}{2}\bar{z}_{j+1}(z_{j+1}-z_{j})-\frac{i\varepsilon}{\hbar}H(\bar{z}_{j+1},z_{j}) \right] \right)\nonumber\\
&=\int_{z_{0},t_{0}}^{z_{f},t_{f}}D[z(t)]D[\bar{z}(t)]\exp\left(\frac{i}{\hbar}\int_{t_{0}}^{t_{f}}dt\,L[\dot{z}(t),z(t),t]\right)~,
\end{flalign}
where
\begin{equation}
L[\dot{z}(t),z(t),t]:=\frac{i\hbar}{2}\left(\bar{z}(t)\dot{z}(t)-\dot{\bar{z}}(t)z(t)\right)-H(\bar{z}(t),z(t))~,
\end{equation}
$d^{2}z=d\textrm{Re}[z]d\textrm{Im}[z]$, $\varepsilon=1/N$, and $H(\bar{z}_{j+1},z_{j})$ is the normal-ordered Hamiltonian after substituting $a^\dagger \rightarrow \bar{z}_{j+1}$ and $a \rightarrow z_j$.
 
We now chose $H=Ea^\dagger a$, which corresponds to $H(\bar{z}(t),z(t))=E|z(t)|^{2}$, and see that for this choice the coherent states path integral in Eq.\ (\ref{cs_path}) is mathematically identical to that in Eq.\ (\ref{Zj}). However, the propagator between two coherent states, given $H=E a^\dagger a$, can also be computed directly from the definition in Eq.\ (\ref{cs_prop}). Using the properties of coherent states we get
\begin{equation}
K(z_{f},z_{0},t)=\langle z_{f}|e^{-\frac{i}{\hbar}Ea^{\dagger}at}|z_{0}\rangle=e^{-\frac{1}{2}(|z_{f}|^{2}+|z_{0}|^{2})}\exp\left(e^{-\frac{i}{\hbar}Et}\bar{z}_{f}\,z_{0}\right)~.
\end{equation}
The path integral in Eq.\ (\ref{Zj}) is then
\begin{equation}
Z_{j}=e^{-\frac{1}{2}(|a_{j}(t_{\rm e})|^{2}+|a_{j}(t_{\rm i})|^{2})}\exp\left(e^{-\frac{i}{\hbar}E_{j}t}\bar{a}_{j}(t_{\rm e})a_{j}(t_{\rm i})\right)~,
\end{equation}
and we arrive at
\beqn
Z&=&e^{-\frac{1}{2}\sum_{j}(|a_{j}(t_{\rm e})|^{2}+|a_{j}(t_{\rm i})|^{2})}\exp\left(\sum_{j}e^{-\frac{i}{\hbar}E_{j}t}\bar{a}_{j}(t_{\rm e})a_{j}(t_{\rm i})\right) \nonumber\\
&=&\exp\left(\langle \Psi_{\rm e} |e^{-\frac{i}{\hbar}Ht}|\Psi_{\rm i} \rangle-1\right)~. \label{fZ}
\eeqn
In the last step we used the fact that the initial and final states are normalized to 1.

This implies that
\begin{equation}
 e^{-2}\leq|Z|\leq1,
\end{equation}
and that $|Z|$ is maximal when  $|\Psi_{\rm e} \rangle = \exp ( -i Ht/\hbar) |\Psi_{\rm i} \rangle$, that is, when the state has the same time-evolution as in normal quantum mechanics. 
It is worth stressing that this path integral reproduces quantum mechanics exactly, without corrections, because by construction the wave-function only takes the most optimal path. 

\section{Extensions}
\label{quant}

Our motivation for introducing this new path integral is not merely to reproduce quantum mechanics. When generalized the right way, this path integral can also provide a starting point to include a collapse-like process for measurements. This becomes possible when we consider a suitable change of the exponent in Eq. (\ref{Z}) because, as we stressed earlier, the individual paths which we integrate over do not need to fulfil the Schr\"odinger equation. 
Furthermore, in contrast to the usual way of defining transition amplitudes from the path integral, the expression Eq.\ (\ref{Z}) is not necessarily linear in the wave-function. It just happens to be linear on the optimal path. 

A natural way to include a collapse-like process is to consider the initial and final state to also include the detector (and possibly other transformation devices), and extend Eq.\ (\ref{Z}) to
 \beqn
Z ( \Psi_{\rm i} \to \Psi_{\rm e} )  = \intsum_{|\Phi\rangle} \exp \left(  - { i} \int {\rm d} t \langle \Phi (t) | {i} \partial_t - \hat{H}/\hbar | \Phi(t) \rangle  - \int {\rm d} t Q (|\Phi(t) \rangle) \right)~, \label{Zq}
\eeqn
where $Q$ is a measure of ``quantumness". One can have in mind here for example the quantum discord \cite{Ollivier:2001fdq} or other measures that quantify how strong the quantum properties of a state are \cite{measures,Hollands:2017dov, Vedral}. 

Intuitively, a state with high quantumness is a macroscopic state with many degrees of freedom that nevertheless has pronounced non-classical features, like for example the paradigmatic Schr\"odinger-cat states or, somewhat more mundanely, superpositions of detector pointer states. For final states with high quantumness, $|Z|$ will be exponentially damped, which means that these states will be very unlikely outcomes of the evolution. On the other hand, a detector in a pointer state has comparably low quantumness and is hence likely to maximize the generating functional.

Such an additional contribution in the path integral should hence make the system prefer final states with low quantumness, such as detector pointer states, in order to maximize $|Z|$. This way, the new path integral could serve to solve the measurement problem without violating locality, and instead by exploiting future-input dependence.  And since the formalism we use is close to that of normal quantum mechanics, the path integral offers a natural framework to be generalized by second quantizations to a quantum field theory.

However, with the measures of quantumness that are commonly used in the literature, evaluating the path integral becomes mathematically intractable. A possible way to proceed would be to use an effective approximation for quantumness that exploits our knowledge that the detector pointer basis exists and thus does not have to be computationally identified from all the degrees of freedom.

\section{Conclusion}
\label{con}
We have shown here that the time-evolution of quantum mechanics can be exactly reproduced from a future-input dependent path integral. We have further argued that this path integral can be extended to also include a collapse-like process for quantum measurements.

\section*{Acknowledgements}

We gratefully acknowledge support from the Fetzer Franklin Fund.

\end{document}